\PassOptionsToPackage{pdfpagelabels=false}{hyperref} 
\expandafter\def\csname ver@fixltx2e.sty\endcsname{}
\documentclass[useAMS,usenatbib]{mnras}

\usepackage{graphicx,epsfig,subfigure}

\usepackage{graphicx}

\usepackage{url}

\usepackage{amsmath}
\usepackage{amssymb}

\usepackage{float}

\title[Dark Matter in ETGs]
   {The quantity of dark matter in early-type galaxies and its relation to the environment}

\author[Nigoche-Netro et al.]{
A. Nigoche-Netro$^1$\thanks{E-mail: anigoche@gmail.com}, G. Ramos-Larios$^1$, P. Lagos$^2$$^,$$^7$, E. de la Fuente$^1$, 
 \newauthor
  A. Ruelas-Mayorga$^5$, J. Mendez-Abreu$^3$$^,$$^4$, S. N. Kemp$^1$, R. J. Diaz$^6$. \\
    $^1$Instituto de Astronom\'ia y Meteorolog\'ia, CUCEI, Universidad de Guadalajara, Guadalajara, Jal. 44130, M\'exico. \\
    $^2$Centre for Space Research, North-West University, Potchefstroom 2520, South Africa. \\
    $^3$Instituto de Astrof\'isica de Canarias, Calle V\'ia L\'actea s/n, E-38205 La Laguna, Tenerife, Spain. \\
    $^4$Departamento de Astrof\'isica, Universidad de La Laguna, E-38206, La Laguna, Tenerife, Spain. \\
$^5$Instituto de Astronom\'ia, Universidad Nacional Aut\'onoma de M\'exico, Cd. Universitaria, M\'exico, D.F. 04510, M\'exico. \\
    $^6$Gemini Observatory, 950 N Cherry Ave, Tucson AZ, USA.\\
    $^7$Instituto de Astrof\'isica e Ci\^encias do Espa\c{c}o, Universidade do Porto, CAUP, Rua das Estrelas, 4150-762 Porto, Portugal.
}

\date{Accepted 2018. Received 2018; in original form 2018}

\pubyear{2018}

\begin{document}
\label{firstpage}
\pagerange{\pageref{firstpage}--\pageref{lastpage}}
\maketitle

\begin{abstract}

 We study the behavior of the dynamical and stellar mass inside the effective radius of early-type galaxies (ETGs) as a function of environment considering Newtonian dynamics, different surface--brightness profiles, different initial mass functions (IMF) and different redshift ranges. We use several samples of ETGs --ranging from 19,000 to 98,000 objects-- from the ninth data release of the Sloan Digital Sky Survey. We assume that any difference between the dynamical and stellar mass is due to dark matter and/or a non-universal IMF. The main results, considering samples in the redshift range 0.0024 $\leq\;z\;\leq$ 0.35 are: (i) the amount of dark matter inside ETGs depends on the environment; (ii) ETGs in low density environments span a wider dark matter range than ETGs in dense environments; (iii) the amount of dark matter inside ETGs in the most dense environments will be less than approximately 55\%--75\% of the dynamical mass; (iv) the accurate value of this upper limit depends on the impact of the IMF on the stellar mass estimation; (v) in the case of an ETGs sample which is approximately complete for log$({\bf M_{Virial}}/{\bf M_{Sun}}) > 10.5$ and in the redshift range 0.04 $\leq\;z\;\leq$ 0.08 we find that the amount of dark matter in the most dense environments will be less than approximately 60\%--65\% of the dynamical mass.

\end{abstract}

\begin{keywords}
 Galaxies: fundamental parameters. Galaxies: photometry, distances and redshifts. Cosmology: dark matter.
\end{keywords}



\section{Introduction}
\label{intro}

The present paper studies the amount of dark matter  inside the effective radius of ETGs as a function of environment, within the hierarchical $\Lambda$ cold dark matter $(\Lambda$CDM$)$ scenario \citep{NAFRWH1996,BULLOCKetal2001,MADUandVA2008}. This scenario tells us that galaxies form inside halos of dark matter (DM). The amount of DM found inside the effective radius $(r_{e})$ of any galaxy may be considered as the central part of the DM halo which was originally at that position. This might have been affected by the galaxy formation process and by gravitational interactions with the matter that goes into forming the central parts of a galaxy. We know that many of the galactic properties we observe today are due to the way in which the halo increases its mass. This process is usually described as the halo's mass aggregation history (MAH; \citealt{MAetal2007}). The MAH leaves its mark on both the halo and the galaxy, for example on the star formation rates (SFRs) and gas infall in disc galaxies \citep{ARandF2000,Muetal2002,vB2002}, and on the morphology of galaxies \citep{Kaetal1993,Baetal1996,SOandPR1999,COetal2000,GRetal2001,SPetal2001,STandNA2002,MAetal2006}. 

There is no doubt that galaxy formation is one of the most pressing issues in modern astrophysics, and though theoretical models of this phenomenon exist (e.g., \citealt{GO1977,vKandFR2011,SOandDA2015} and references therein), observational data help to establish a plausible set of scenarios in  which galaxies form. Surveys such as the Sloan Digital Sky Survey (SDSS; e.g., \citealt{ABetal2003,ABetal2009,ADetal2008}), and the UKIRT Infrared Deep Sky Survey (UKIDSS; \citealt{law07}) have provided many high-quality observational data sets that have permitted the better understanding of individual galactic properties, as well as their properties as a class of entities, such as the scaling relations (Fundamental Plane, Kormendy, Faber--Jackson and Tully-Fisher Relations; e.g., \citealt{DJandDA1987,DRetal1987,kor77,fab76,tul77}).

The dependence of galaxy properties on their surrounding environment has been addressed in many papers (e.g., \citealt{DR1980,POandGE1984,WHetal1993,DOetal2002,GOTetal2003,nig07}), in particular, the dependence of Star Formation Rate (SFRs) and colours (e.g., \citealt{GOMetal2003,BLetal2003,HOetal2004,KAetal2004,TAetal2004,CRetal2005} and references therein). One of the main correlations found in these studies is that the morphological mix of spirals and ellipticals (meaning the relative numbers of different types of galaxies) changes depending on the environment's density.

Many studies have concentrated on ETGs since they present a number of properties (old stellar population, low gas fraction, and no or very little star formation) that makes them a homogeneous group suitable for studying their evolution as a function of time, or equivalently as a function of redshift $(z)$. ETGs obey the Fundamental Plane Relation \citep{DJandDA1987,DRetal1987}, for which the coefficients determined from observations differ from those expected theoretically. This result has been interpreted by \citet{DRetal1987} as a possible variation of the mass to light (M/L) ratio as a function of total luminosity and/or total mass. This variation may be caused by a different amount of DM content in each early-type galaxy, therefore, many research groups have attempted to measure the DM contents of ETGs (see \citealt{cio96,BUetal1997,GRandCO1997,PRandSI1997,TRBUandBE2004,cap06,DOetal2006,GR2009,TOetal2009,LBetal2010b,tor12,nig15,nig16,tor18} among others). The DM content of the central parts of ETGs (within $r_{e}$) have been measured for nearby samples using various techniques (\citealt{PAetal2014,cap06,hyd09a,TOetal2009,NAROandTO2010,tor12b,tor18}), and for samples at intermediate redshift using the technique of gravitational lensing (\citealt{CAetal2009,AUetal2010b,CAandTO2010,TOetal2010,FAetal2011,MOetal2011, pos15}).

Observational estimates of the amount of DM within $r_{e}$ appear to indicate that DM haloes are cuspy \citep{THetal2009,NAROandTO2010} and, independently of the galaxy mass distribution, the DM amount tends to increase with velocity dispersion, size, stellar mass, and luminosity of the galaxy \citep{FESAandWI2005,cap06,FOetal2008,TOetal2009,NAROandTO2010,AUetal2010b,LEetal2011,TOetal2010,nig16}, in agreement with theoretical predictions of the DM fraction $(f_{DM})$ in central regions of galaxies \citep{RUandSP2009,HINAOS2013,WUetal2014}.

In contrast with the results of \cite{THetal2009} and \cite{NAROandTO2010}, \citet{TRBUandBE2004} reached the conclusion that the amount of DM within $r_{e}$ does not depend on the galaxy properties. However,  the total stellar-to-dark matter mass ratio depends strongly on the mass of the galaxy and also may be connected to the total star formation efficiency \citep{BEetal2000,MAandHU2002,NAetal2005,MANetal2006,VAetal2007,COandWE2009,MOetal2010,ALetal2016}. There are, however, anticorrelations of $f_{DM}$ with mass that have been found by \citet{GRetal2009}, \citet{GR2010}, and \citet{GRandGO2010}. These results may indicate that the amount of DM within $r_{e}$ correlates with the baryonic mass of a galaxy, however this apparent $f_{DM}$ trend with mass could disappear completely if a non-universal IMF is used \citep{tho11,tor13,nig16}. It is important to mention that the IMF is the most important source of uncertainty in the calculation of the relative amounts of stellar and DM mass, in the central regions of a galaxy (\citealt{tor18}).

\citet{tor18} studied the $f_{DM}$ within $r_{e}$ of a sample of 3800 well-observed galaxies from the Kilo Degree Survey (KiDS) up to $z \sim 0.65$. They found that $f_{DM}$ (within $r_{e}$) correlated with structural parameters, mass, and central stellar density indicators, and showed that at larger redshifts (up to $z \sim 0.65$) the local correlations were still observed. At fixed total stellar mass ({\bf M$_{\bf Star}$}), the amount of DM seems to be smaller for galaxies at higher redshifts as mentioned previously.

\citet{tor12b} performed an analysis of the amount of DM inside $r_{e}$ in ETGs. They used approximately 4500 ETGs from the SDSS and supplemented their data with $YJHK$ photometry from the DR2 of the UKIDSS-Large Area Survey. They found that $f_{DM}$ increased steeply as a function of $r_{e}$, S\'ersic index, velocity dispersion $(\sigma)$, stellar mass, and dynamical mass. Galaxies with denser stellar cores appeared to have smaller amounts of DM inside  $r_{e}$. They also found that the central DM content of ETGs did not depend significantly on the environment where galaxies resided, with group and field ETGs having similar DM trends. On the other hand, \citet{cor17} analysing a small sample of ETGs in low density environments found that they have a lower content of halo dark matter with respect to ETGs in high-density environments.

\citet{nig15} studied several samples of relatively nearby ETGs $(0.0024 < z < 0.3500)$ taken from the SDSS-DR9. They measured the amount of dynamical mass (${\bf M_{Virial}}$) and stellar mass (${\bf M_{Star}}$) within one $r_{e}$ considering Newtonian dynamics, virial equilibrium, different surface-brightness profiles, different IMFs, and different redshifts. They established that any difference between ${\bf M_{Virial}}$ and ${\bf M_{Star}}$ was due to the presence of DM, or a non-universal IMF, or both. They found that using linear fits to the distribution of galaxies in the plane log$({\bf M_{Virial}/{\bf M_{Sun}}})$ - log$({\bf M_{Star}/{\bf M_{Sun}}})$ in order to obtain an estimation of the mean values of the DM was not a reliable method because this distribution had a high intrinsic dispersion and this intrinsic dispersion could depend on several variables such as mass, wavelength, environment and/or redshift. An approximation to the dependence of the DM on the above-mentioned variables was made in \citet{nig16}. They found that ETGs cover a wide range of the amount of DM inside them, which goes from almost 0 to approximately 70 per cent of the dynamical mass, with the mass and redshift responsible for this behaviour. Besides, their results indicated that the amount of DM increased as a function of dynamical mass and decreased as a function of redshift. The latter result is in agreement with the work of \cite{BEIFIORIetal2014}, \cite{tor14} and \cite{tor18} who found that high-$z$ ETGs have smaller amounts of DM than the local ETGs.

We can see that there are only a few papers in the literature that address the question of the DM inside  $r_e$ of ETGs as a function of the environment. Therefore, in the present paper we will address the behaviour of the DM as a function of the environment where the galaxies reside. We will also compare our results with those few papers where this issue is analysed.

It is important to note that in this paper the difference between dynamical and stellar mass, defined as log$({\bf M_{Virial}/{\bf M_{Sun}}})$ - log$({\bf M_{Star}/{\bf M_{Sun}}})$, was obtained considering an universal (constant) IMF (see section \ref{sec:stellar&virial}). This difference between dynamical and stellar mass can be used to calculate the DM fraction --$f_{DM}$-- and it may be used to compare with other results from the literature, if and only if, an universal IMF is used in the calculation of the $f_{DM}$.

This paper is organised as follows. In section \ref{sec:sample} we present our sample of galaxies and the method by which it was selected. In section \ref{sec:stellar&virial} we calculate the stellar and virial masses of the galaxies in our sample. Section \ref{sec:systematics} analyses possible systematic errors. In section \ref{sec:discussion} we discuss our findings and finally in section \ref{sec:conclusions} we present our conclusions.

\section{The sample of ETGs}
\label{sec:sample}

 We use a sample of approximately 98000 ETGs from SDSS-DR9 \citep{yor00} in the $g$ and $r$ filters. The galaxies are distributed over a redshift interval $0.0024 < z < 0.3500$. This sample will be called hereafter, the ``Total--SDSS--Sample". The selection criteria are the same we used in earlier papers, for full details see \citet{hyd09a}, and \citet{nig10}. 

Selecting only ETGs from the morphological classification of the Zoospec catalogue \citep{lin08} and considering our selection criteria the sample is reduced to approximately 27,000 ETGs. The galaxies with this added criterion have a higher probability of being ETGs. This sample shall be referred to as ``Morphological--SDSS--Sample". In addition, if we want to control possible streaming motions, redshift bias, and evolutionary effects, we have to compile a relatively nearby and volume-limited sample. The redshift range 0.04 $\leq\;z\;\leq$ 0.08 corresponds to a volume that fits these characteristics \citep{nig08,nig09}. The resulting sample contains approximately 19 000 ETGs. This sample is approximately complete for log$({\bf M_{Virial}/{\bf M_{Sun}}}) > 10.5$ \citep{nig10,nig11}. We shall refer to it as the ``Homogeneous--SDSS--Sample".

The photometry and spectroscopy of the galaxy samples drawn from the DR9 have been corrected for different biases and are the same that we have used in previous papers, see \citet{nig15} for full details.

\section{The stellar and virial mass of the ETGs}
\label{sec:stellar&virial}

\subsection{The stellar mass}

The total stellar or luminous mass was obtained considering different stellar population synthesis models, an universal IMF (Salpeter or Kroupa), and different surface--brightness profiles (de Vaucouleurs or S\'ersic; see \citealt{nig15} for details). These methods result in three mass estimations.

\begin{itemize}

\item 	de Vaucouleurs Salpeter--IMF stellar mass. This mass was obtained considering a de Vaucouleurs profile and an equation for stellar mass--to--light (M/L) ratios (\citealt{bel03}) obtained from fits of optical and near--infrared galaxy data with simple stellar population synthesis models and a ``diet" Salpeter IMF. The used ``diet" Salpeter stellar IMF has the same luminosity and colours as a normal Salpeter IMF, but with only 70\% of the mass due to a lower number of faint low-mass stars. We used the following equation from \cite{bel03}:

\begin{equation}  
\label{eq:eq1}
{\bf \frac{M}{L} } = 10^{a_{g} + b_{g} (\it M_{g}-M_{r})},
\end{equation}

where ${\bf L}$ is the luminosity and ${\bf M}$ is the stellar mass, $M_{g}$ and $M_{r}$ are the magnitudes in the $g$ and $r$ filters, respectively, and $a_{g}$ and $b_{g}$ are scale factors.

 \item	S\'ersic Salpeter--IMF stellar mass. This mass was obtained considering S\'ersic parameters and a Salpeter IMF using equation \ref{eq:eq1}. The S\'ersic parameters were obtained from Petrosian parameters using the methodology of \cite{gra05}

\item	Kroupa--IMF stellar mass. This mass was calculated by the MPA--JHU group \citep{kau03,bri04,trm04} considering model magnitudes, a universal Kroupa IMF, Bayesian methodology, and model grids described in \cite{kau03}. These stellar masses were obtained from the MPA--JHU site\footnote{$http://www.mpa-garching.mpg.de/SDSS/$}. These masses also can be obtained from the GALSPEC catalogue of the SDSS--DR9 \footnote{$http://www.sdss3.org/dr9/algorithms/galaxy_mpa_jhu.php$}.

\end{itemize}

It is important to mention that these masses are total masses. Due to the fact that our goal is to analyze only the mass inside $r_{e}$ we take into account that within a sphere of radius  $r_{e}$ only contains 42\% of the total stellar mass  \citep{sch10}.

\subsection{The dynamical mass}
\label{sec:sec3.2}

The total dynamical or virial mass was obtained using an equation from \citet{pov58}. This method assumes Newtonian mechanics and virial equilibrium for the galaxies in question. The equation is as follows:

\begin{equation}  \label{eq:eq2}
{\bf M_{Virial}} \sim  K(n) \frac{ r_{e} \sigma_{e}^{2}}{G},
\end{equation}

where the variables ${\bf M_{Virial}}$, $r_{e}$ and, $\sigma_{e}$ represent the total virial mass, the effective radius and the velocity dispersion inside $r_{e}$, respectively. $G$ is the gravitational constant and $K(n)$ is a scale factor that depends on the S\'ersic {\it index} (n) as follows \citep{cap06}:

\begin{equation}  \label{eq:eq3}
 K(n) = 8.87 - 0.831 n + 0.0241 n^{2},
\end{equation}

Equation \ref{eq:eq2} considers an idealised situation and does not take into account possible effects on the dynamical mass estimations of ETGs due to environment, shape, rotation and velocity dispersion anisotropies. In sections \ref{sec:systematics} and \ref{sec:discussion} we will discuss these effects. 

The amount of mass within an effective radius also corresponds to 0.42 times the value calculated from equation \ref{eq:eq2}. This mass may or may not be luminous.

Here we should mention that the estimations of the dynamical mass in previous works by \citet[][]{nig15,nig16} were made considering a scale factor K(4) = 5.95 (de Vaucouleurs profile) for all galaxies. The average difference between this estimation of dynamical mass and that found when the S\'ersic profile (equation \ref{eq:eq3}) is used is approximately 0.05 dex towards higher mass.

Some papers claim that the IMF used in the stellar mass estimation is not universal but instead depends on the velocity dispersion and mass of galaxies \citep{cap12,dut13,ber18}. Since the current correction methods to the stellar mass found in the literature use the virial mass or the velocity dispersion, the stellar mass corrected with such methods will be correlated with our estimation of virial mass (equation \ref{eq:eq2}). Because of this, we do not correct our stellar mass by a non-universal IMF. We  have to take this into account in our results and we discuss it in section \ref{sec:discussion}.

\subsection{Density of galaxies}
\label{sec:sec3.3}

The projected density of galaxies was computed following the method described in \citet{ague09}. They used the projected co-moving distance to the Nth nearest neighbour $(d_{N})$ of the target galaxy as follows:

\begin{equation}  \label{eq:eq4}
{ \Sigma_{N}} \sim \frac{ N }{\pi (d_{N})^{2}},
\end{equation}

The nearest neighbours were calculated using two different samples which are defined as follows:

    i) Spectroscopic sample. This sample was selected considering only those galaxies with spectroscopic redshift in a velocity range of $\pm$1000 km s$^{-1}$ from the target galaxy and within a magnitude range of $\pm$2 mag. These two constraints are similar to those used by \citet{balo04} and allow us to limit the contamination by background/foreground galaxies even if we are working with projected distances.

    ii) Photometric sample. This sample was selected considering only those galaxies with photometric redshifts within $\pm$0.1 of the target galaxy \citep[see][]{bald06}. This range approximately corresponds to the typical photometric redshift error in SDSS. We also imposed the magnitude limitation of $\pm$2 mag as for the spectroscopic sample.
 
The photometric sample is expected to be more complete than the spectroscopic one, however, the large uncertainties in the photometric redshifts make this sample more prone to contamination. Therefore, the density of galaxies derived from either the spectroscopic or photometric sample can be considered as a lower or upper limit to the actual density, respectively. In addition, to account for possible edge effects in our sample, we discard those galaxies with $d_{N}$ greater than the distance to the edge of the survey, as these galaxies will have much more uncertain environmental densities.

The density of galaxies was calculated using the third, fifth, eighth, and tenth nearest neighbours, for both the spectroscopic and photometric samples.

\begin{figure}
   \begin{center}

 \includegraphics[angle=0,width=0.95\columnwidth]{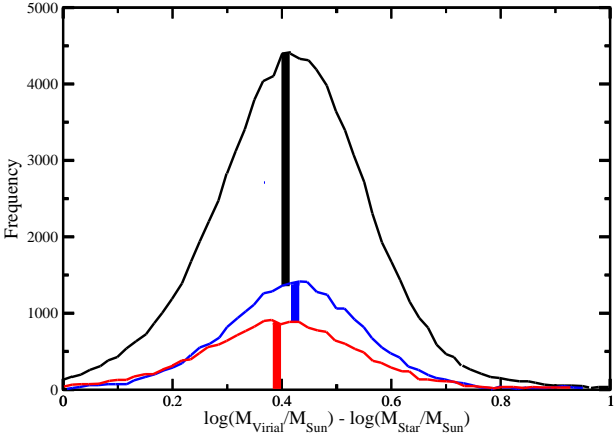}
   
  \caption{Frequency distributions of the logarithmic difference between dynamical and stellar mass inside $r_{e}$ for different samples considering a Kroupa-IMF stellar mass. Black, blue and red histograms correspond to the total, morphological, and homogeneous SDSS samples, respectively. The median values (shaded columns) of log$({\bf M_{Virial}/{\bf M_{Sun}}})$ -- log$({\bf M_{Star}/{\bf M_{Sun}}})$ for each of the distributions are 0.416, 0.421 and 0.391, respectively.}
  
\label{FigVibStab1}
         
 \end{center}

   \end{figure}

\begin{figure}
   \begin{center}

\includegraphics[angle=0,width=0.95\columnwidth]{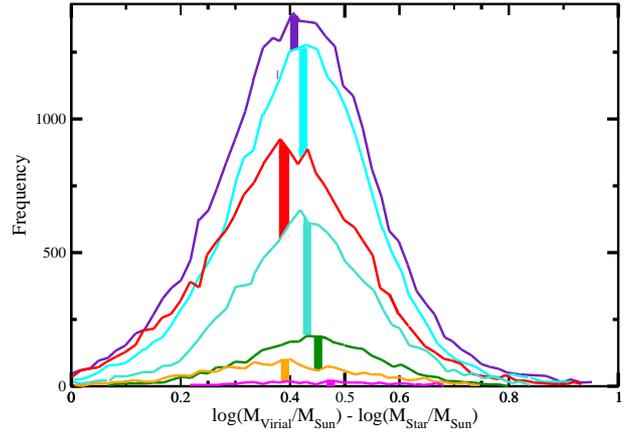}

         \caption{Frequency distribution of the logarithmic difference between dynamical and stellar mass inside $r_{e}$ for the Total-SDSS sample (Kroupa--IMF stellar mass) at different redshifts. Each color represents a different redshift range as follows: orange 0.00 $\leq\;z\;\leq$ 0.04, red 0.04 $\leq\;z\;\leq$ 0.08, indigo 0.08 $\leq\;z\;\leq$ 0.12, cyan 0.12 $\leq\;z\;\leq$ 0.16, turquoise 0.16 $\leq\;z\;\leq$ 0.20, green 0.20 $\leq\;z\;\leq$ 0.24  and magenta 0.24 $\leq\;z\;\leq$ 0.28. The median values of the the difference between log dynamical and stellar mass for each one of the distributions are 0.388, 0.391, 0.411, 0.427, 0.434, 0.450 and 0.481, respectively.}

         \label{FigVibStab2}
         \end{center}
   \end{figure}

\subsection{Frequency distribution of the difference between dynamical and stellar mass}
\label{sec:sec3.4}

Fig. \ref{FigVibStab1} shows the frequency distributions of the logarithmic difference between the dynamical and stellar mass for different samples considering the Kroupa--IMF stellar mass. Black, blue, and red histograms correspond to the total--SDSS-sample, morphological--SDSS--sample, and homogeneous--SDSS--sample, respectively. In this figure we can see that the maximum of the frequency distribution of the different samples is found between 0.35 -- 0.45 in log$({\bf M_{Virial}/{\bf M_{Sun}}})$ -- log$({\bf M_{Star}/{\bf M_{Sun}}})$. The median values of log$({\bf M_{Virial}/{\bf M_{Sun}}})$ -- log$({\bf M_{Star}/{\bf M_{Sun}}})$ for the total, morphological and homogeneous sample respectively are 0.416, 0.421 and 0.391. If we consider the differences caused by the factor K in the estimation of the dynamical mass (see section \ref{sec:sec3.2}), the previous results are in agreement with those of \citet[][see their section 5 and table 1]{nig16}. It is important to mention that the average values reported in table 5 by \citet{nig16} and those found here, have been obtained without considering that the difference between dynamical and stellar mass depends on mass \citep{tor12b,cap13,nig16} and redshift \citep{,nig16}. In the following we will analyse the above-mentioned dependencies.

A first approximation to the behaviour of the frequency distribution of log$({\bf M_{Virial}/{\bf M_{Sun}}})$ -- log$({\bf M_{Star}/{\bf M_{Sun}}})$ at different redshifts considering the Kroupa-IMF stellar mass is shown in Fig. \ref{FigVibStab2}. In this figure we have samples in redshift intervals of width 0.04, starting at 0.00 $\leq\;z\;\leq$ 0.04 and finishing at 0.24 $\leq\;z\;\leq$ 0.28. The median values of log$({\bf M_{Virial}/{\bf M_{Sun}}})$ -- log$({\bf M_{Star}/{\bf M_{Sun}}})$ for 0.00 $\leq\;z\;\leq$ 0.04 (orange), 0.04 $\leq\;z\;\leq$ 0.08 (red), 0.08 $\leq\;z\;\leq$ 0.12 (indigo), 0.012 $\leq\;z\;\leq$ 0.16 (cyan), 0.16 $\leq\;z\;\leq$ 0.20 (turquoise), 0.20 $\leq\;z\;\leq$ 0.24 (green), and 0.24 $\leq\;z\;\leq$ 0.28 (magenta) are 0.388, 0.391, 0.411, 0.427, 0.434, 0.450, and 0.481 respectively. In Fig. \ref{FigVibStab2} we can see that the maxima of the distribution of samples in different redshift ranges cover the interval 0.38 - 0.48 of the difference between log virial and stellar mass. Indeed we can see a mild shift of the maximum at large redshift, as the maximum is close to 0.4 for nearby galaxies and close to 0.5 for distant galaxies. This result seems to be the opposite to that found in \citet{nig16} and by other authors, where they found that the difference between virial and stellar mass decreases as a function of redshift. The most plausible explanation of this apparent contradiction is that the dependence of log$({\bf M_{Virial}/{\bf M_{Sun}}})$ -- log$({\bf M_{Star}/{\bf M_{Sun}}})$ on mass, reported by different authors, has not been taken into account in the previous analysis. Therefore, since the difference between dynamical and stellar mass increases as a function of mass \citep{tor12b,cap13,nig16} and considering that the samples are affected by Malmquist bias (samples at greater redshift contain only massive galaxies), the maximum of this difference between masses is shifted at higher redshift. To deal with this effect, we have to analyse complete samples with respect to mass. Due to the impossibility of having a complete sample for the entire mass range, we can use relatively nearby samples containing all the massive galaxies and a large quantity of less massive galaxies. The samples that match the above-mentioned characteristics are those in the redshift range 0.00 $\leq\;z\;\leq$ 0.04 and 0.04 $\leq\;z\;\leq$ 0.08 (homogeneous--SDSS sample). Analysing these two samples we found that the maxima are located at 0.388 and 0.391, respectively. However the sample contained in the redshift range 0.00 $\leq\;z\;\leq$ 0.04 could be affected by streaming motions, so to avoid this possible bias we will use the result of the homogeneous--SDSS sample.

\begin{figure}
   \begin{center}

\includegraphics[angle=0,width=0.95\columnwidth]{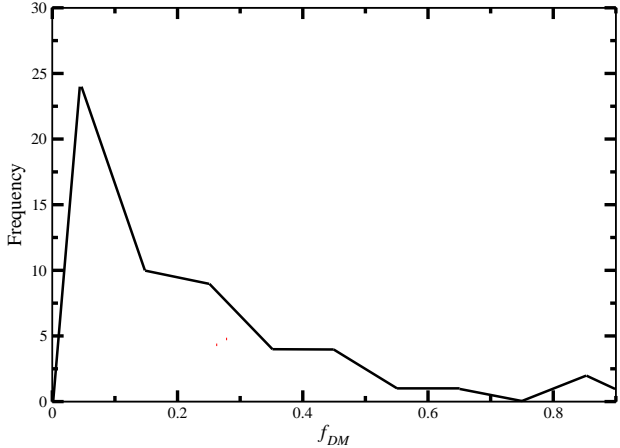}

         \caption{Frequency distribution of the dark matter fraction $(f_{DM})$ inside $r_{e}$ of the 55 massive lens ETGs from \citet{pos15}.  }

         \label{FigVibStab3}
         \end{center}
   \end{figure}

\citet{pos15} studied 55 massive lensed ETGs ($10.5 < log({\bf M_{Virial}/{\bf M_{Sun}}}) < 11.7$)  from the Sloan Lens Advanced Camera for Surveys (SLACS) in the redshift range 0.06 $\leq\;z\;\leq$ 0.36. They found that $f_{DM}$ inside the effective radius of their ETGs is relatively low. From table 3 of \citet{pos15}, the mean $f_{DM}$ is approximately 0.2. The distribution of the DM fraction can be seen in Fig. \ref{FigVibStab3}. The maximum of the distribution of dark matter fraction is close to zero. These results appear to be different to those found here. The difference could be due to systematics (only massive galaxies, non--homogeneous redshift sample, low number of galaxies) that may affect \citet{pos15} sample and also that our stellar mass estimation was calculated using a universal IMF which may not be appropriate for the entire range of masses of our samples (see section \ref{sec:discussion} for more details).

   \begin{figure*}
   \begin{center}

      \includegraphics[angle=0,width=16cm]{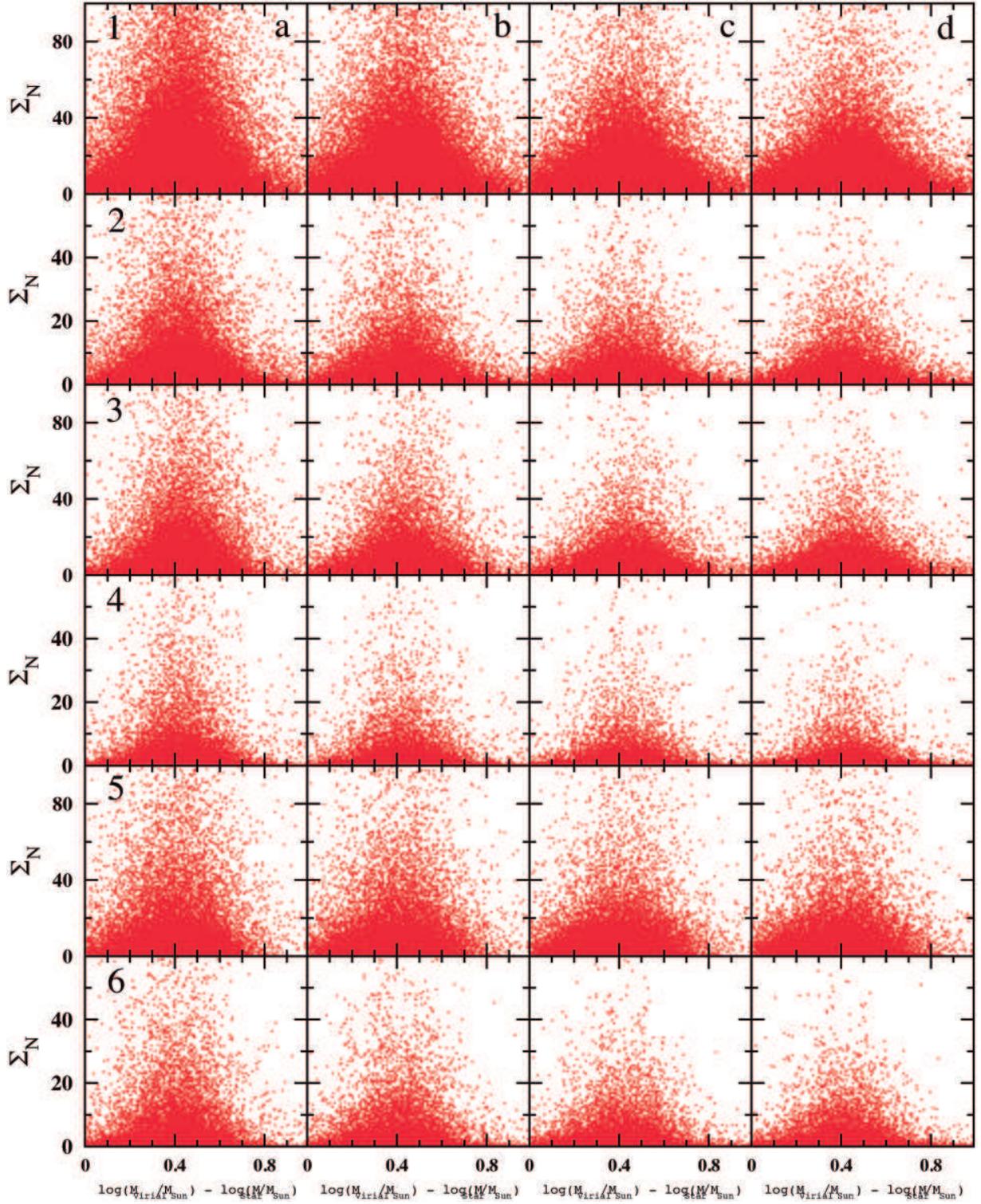}
      
         \caption{Distribution of the logarithmic difference between dynamical and stellar mass inside $r_{e}$ as function of density of galaxies for different samples of ETGs considering a Kroupa-IMF stellar mass. Rows 1-2, 3-4 and 5-6 correspond to the total, morphological, and homogeneous SDSS samples, respectively. Rows 1, 3, and 5 are data from the photometric sample. Rows 2, 4 and 6 are data from the spectroscopic sample. Columns a, b, c, d are the data considering the third, fifth, eight, and tenth nearest neighbours respectively.}

         \label{FigVibStab4}
         \end{center}
   \end{figure*}
 
\subsection{Distribution of the difference between dynamical and stellar mass as a function of density of galaxies considering a Kroupa-IMF stellar mass}
\label{sec:sec3.5}

Fig. \ref{FigVibStab4} shows a mosaic of the logarithmic difference between dynamical and stellar mass as a function of density of galaxies for different samples considering only the Kroupa--IMF stellar mass data. Rows 1--2, 3--4, and 5--6 correspond to the total, morphological, and homogeneous SDSS samples, respectively. Rows 1, 3, and 5 are data from the photometric sample and rows 2, 4, and 6 are data from the spectroscopic sample. Columns a, b, c, d are the data considering the third, fifth, eighth, and tenth nearest neighbours, respectively. From Fig. \ref{FigVibStab4} we can see that the distribution of galaxies is relatively similar for all samples, indicating that our results are not dependent on which galaxy sample is used or which galaxy is considered the nearest neighbour. We can see that the distribution is not random but rather has a bell--shape. It is interesting to note that galaxies in the lowest density region cover the whole range of difference between virial and stellar mass. This range of the difference between masses decreases while the density increases. In all cases we can see that the maximum of the density distribution is at about 0.35 -- 0.45 in log$({\bf M_{Virial}/{\bf M_{Sun}}})$ -- log$({\bf M_{Star}/{\bf M_{Sun}}})$. If we translate these results to linear values we find that the maximum of the density distribution has a difference between masses of about 55\% - 65\% of the virial mass.

The previous result has been obtained considering the dependence of the scale factor K on the S\'ersic brightness profile (n; see equation \ref{eq:eq3}), but if we use K = 5.9 (de Vaucouleurs profile) for the galaxies in all samples the results are similar. The most conspicuous difference between these two procedures is a small shift (approximately 0.05 dex) towards higher mass difference of the samples using K(n).

\begin{figure}
   \begin{center}

       \includegraphics[angle=0,width=0.95\columnwidth]{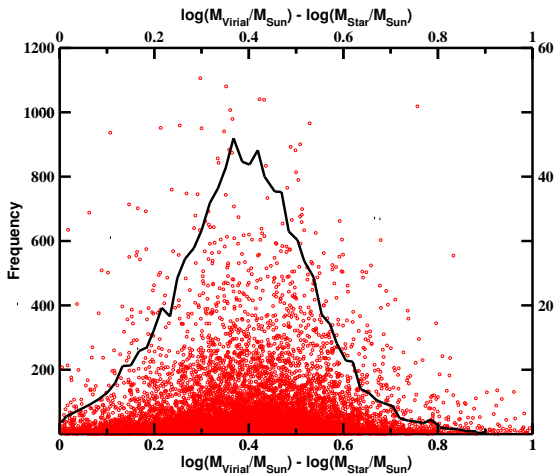}

         \caption{Density of galaxies (red) and frequency distribution (black) of the logarithmic difference between dynamical and stellar mass inside $r_{e}$ for the spectroscopic--homogeneous--SDSS sample (Kroupa--IMF stellar mass) considering the tenth nearest neighbour.}

         \label{FigVibStab5}
         \end{center}
   \end{figure}

The results from Fig. \ref{FigVibStab4} appear to be similar to the results of the frequency distribution of the different samples analysed in section \ref{sec:sec3.4}. The maximum of the density distribution as function of the difference between virial and stellar mass appears to be similar to the maximum of the frequency distribution of the difference between virial and stellar mass. This similarity could be due to the fact that there are more galaxies with the log$({\bf M_{Virial}/{\bf M_{Sun}}})$ -- log$({\bf M_{Star}/{\bf M_{Sun}}})$ close to 0.4, and then the probability that those galaxies are in dense environments is higher. In Fig. \ref{FigVibStab5} we can see the behaviour previously described. Each red point represents a galaxy in the plane log$({\bf M_{Virial}/{\bf M_{Sun}}})$ -- log$({\bf M_{Star}/{\bf M_{Sun}}})$ vs. density of galaxies for the spectroscopic--homogeneous--SDSS sample considering the tenth nearest neighbour. The black line represents the frequency distribution of log$({\bf M_{Virial}/{\bf M_{Sun}}})$ -- log$({\bf M_{Star}/{\bf M_{Sun}}})$.  We can see that the maximum of the frequency distribution follows approximately the maximum of the density distribution.

\begin{table*}

\renewcommand{\footnoterule}{}  

\caption{Mean values of the difference between log$({\bf M_{virial}/{\bf M_{Sun}}})$ and log$({\bf M_{*}/{\bf M_{Sun}}})$ inside $r_{e}$ for field, loose groups and compact groups/clusters considering several samples of galaxies (Kroupa--IMF stellar mass) and different values for the nearest neighbour (see section \ref{sec:sec3.3}).}


\scalebox{0.45}{
\begin{tabular}{|l|l|c|c|c|c|}

\hline
\hline
                           &                             &                 &                                                      & &  \\
   &&&       {\Huge Mean logarithmic difference between virial and stellar mass}   &  \\
                           &                             &                 &{\Huge  }         &{\Huge}        &  \\
\hline
\hline
                           &                             &                 &                                         &               &  \\              & & {\Huge Field} &{\Huge Loose groups}                          & {\Huge Clusters }   \\
                           &                             &                 &                                                          \\
\hline

        &                             &                 &                                                         \\

{\Huge Total sample}        &   & &    &               \\
                           &                             &                 &                                                         \\
\hline
                           &                             &                 &       &                                                   \\

                                                  & {\Huge Photometric data}        &   & &                   \\
                           &                             &                 &                                                         \\
\hline
                           &                             &                 &         &                                                 \\
 &{\Huge Third nearest neighbour}  &{\Huge 0.406} &{\Huge 0.416} & {\Huge 0.410} \\

&{\Huge Fifth nearest neighbour}   &{\Huge 0.408} & {\Huge 0.416} & {\Huge 0.417}\\

&{\Huge Eighth nearest neighbour}  &{\Huge 0.409} &{\Huge 0.416} & {\Huge 0.411} \\

&{\Huge Tenth nearest neighbour} & {\Huge 0.410} &{\Huge 0.416} & {\Huge 0.425} \\
                           &                             &                 &                                                        \\

\hline
                           &                             &                 &                                         &               &  \\

                        & {\Huge Spectroscopic data}        &   & &                   \\
                           &                             &                 &                                                         \\
\hline

     &                             &                 &                                                        \\

 &{\Huge Third nearest neighbour}  &{\Huge 0.414} &{\Huge 0.420} & {\Huge 0.425} \\

&{\Huge Fifth nearest neighbour}   &{\Huge 0.414} & {\Huge 0.420} & {\Huge 0.411}\\

&{\Huge Eighth nearest neighbour}  &{\Huge 0.415} &{\Huge 0.419} & {\Huge 0.404} \\

&{\Huge Tenth nearest neighbour} & {\Huge 0.406} &{\Huge 0.417} & {\Huge 0.413} \\
                           &                             &                 &                                                        \\

\hline

&   & &    &               \\
{\Huge Morphological sample}        &   & &    &               \\
                           &                             &                 &                                                         \\
\hline
                           &                             &                 &       &                                                   \\

                                                  & {\Huge Photometric data}        &   & &                   \\
                           &                             &                 &                                                         \\
\hline
                           &                             &                 &         &                                                 \\
 &{\Huge Third nearest neighbour}  &{\Huge 0.399} &{\Huge 0.424} & {\Huge 0.434} \\

&{\Huge Fifth nearest neighbour}   &{\Huge 0.400} & {\Huge 0.424} & {\Huge 0.410}\\

&{\Huge Eighth nearest neighbour}  &{\Huge 0.403} &{\Huge 0.424} & {\Huge 0.417} \\

&{\Huge Tenth nearest neighbour} & {\Huge 0.404} &{\Huge 0.424} & {\Huge 0.439} \\
                           &                             &                 &                                                        \\

\hline
                           &                             &                 &                                         &               &  \\

                        & {\Huge Spectroscopic data}        &   & &                   \\
                           &                             &                 &                                                         \\
\hline

     &                             &                 &                                                        \\

 &{\Huge Third nearest neighbour}  &{\Huge 0.418} &{\Huge 0.427} & {\Huge 0.420} \\

&{\Huge Fifth nearest neighbour}   &{\Huge 0.419} & {\Huge 0.428} & {\Huge 0.424}\\

&{\Huge Eighth nearest neighbour}  &{\Huge 0.419} &{\Huge 0.429} & {\Huge 0.421} \\

&{\Huge Tenth nearest neighbour} & {\Huge 0.420} &{\Huge 0.428} & {\Huge 0.393} \\
                           &                             &                 &                                                        \\

\hline

&   & &    &               \\
{\Huge Homogeneous sample}        &   & &    &               \\
                           &                             &                 &                                                         \\
\hline
                           &                             &                 &       &                                                   \\

                                                  & {\Huge Photometric data}        &   & &                   \\
                           &                             &                 &                                                         \\
\hline
                           &                             &                 &         &                                                 \\
 &{\Huge Third nearest neighbour}  &{\Huge 0.383} &{\Huge 0.392} & {\Huge 0.393} \\

&{\Huge Fifth nearest neighbour}   &{\Huge 0.387} & {\Huge 0.392} & {\Huge 0.377}\\

&{\Huge Eighth nearest neighbour}  &{\Huge 0.387} &{\Huge 0.392} & {\Huge 0.397} \\

&{\Huge Tenth nearest neighbour} & {\Huge 0.388} &{\Huge 0.392} & {\Huge 0.398} \\
                           &                             &                 &                                                        \\

\hline
                           &                             &                 &                                         &               &  \\

                        & {\Huge Spectroscopic data}        &   & &                   \\
                           &                             &                 &                                                         \\
\hline

     &                             &                 &                                                        \\

 &{\Huge Third nearest neighbour}  &{\Huge 0.380} &{\Huge 0.402} & {\Huge 0.410} \\

&{\Huge Fifth nearest neighbour}   &{\Huge 0.381} & {\Huge 0.403} & {\Huge 0.387}\\

&{\Huge Eighth nearest neighbour}  &{\Huge 0.383} &{\Huge 0.404} & {\Huge 0.409} \\

&{\Huge Tenth nearest neighbour} & {\Huge 0.384} &{\Huge 0.404} & {\Huge 0.382} \\
                           &                             &                 &                                                        \\

\hline

\end{tabular}}
\label{tab1}
\end{table*}

A different way of making comparisons of structural properties of galaxies is to make samples with different (environmental) densities, for example, \cite{ague09} consider very low-density environments (field)  ($\Sigma_{5} < 1$ Mpc$^{-2}$), loose groups ($\Sigma_{5} > 1$ Mpc$^{-2}$) and compact galaxy groups/clusters ($\Sigma_{5} \sim 10$ Mpc$^{-2}$) to investigate if the environment plays a role in the fraction of barred galaxies. Using these considerations they find that the fraction of barred galaxies does not depend on the environment.

Following the environmental definition of \cite{ague09} we investigate here the possible effects of the environment on the virial and stellar mass difference inside ETGs. We calculate the mean values of the logarithmic mass difference for $\Sigma_{N} < 1$ Mpc$^{-2}$ (field), $\Sigma_{N} > 1$ Mpc$^{-2}$ (loose groups) and 9.9 Mpc$^{-2} < \Sigma_{N} < 10.1$ Mpc$^{-2}$ (compact groups/clusters) where N is the third, fifth, eighth, and tenth nearest neighbour. In Tab. \ref{tab1} we show the results for the total, morphological and homogeneous SDSS samples considering both the photometric and spectroscopic data (see section \ref{sec:sec3.3}).

From Tab. \ref{tab1} we can see that the mean value of the mass difference for the field, loose and compact clusters within the same sample has a variation less than 0.03 dex which, considering the associated errors (see section \ref{sec:systematics}), is not significant. This result agrees with other papers in the literature \citep{tor12b} where they find that there are no differences in dark matter content inside ETGs due to the environment. However it appears to be opposed to the results of the visual inspection of the distribution that was described in previous paragraphs. The explanation for this contradiction can be reasoned as follows: the distribution of galaxies in the log$({\bf M_{Virial}/{\bf M_{Sun}}})$ -- log$({\bf M_{Star}/{\bf M_{Sun}}})$ vs. density plane (see Fig. \ref{FigVibStab4}) for all the samples appears to be symmetric but non-homogeneous (galaxies do not populate the plane uniformly), these properties cause that the distribution inside each one of the areas limited by $\Sigma_{N} < 1$ Mpc$^{-2}$, $\Sigma_{N} > 1$ Mpc$^{-2}$ and 9.9 Mpc$^{-2} < \Sigma_{N} < 10.1$ Mpc$^{-2}$ is also symmetric but non-homogeneous. Given the symmetry of the distributions, the mean values (and also any other measure of the central tendency such as median or mode) of those distributions have to be similar. Therefore, the results for the mean values for the field, loose, and compact groups of galaxies must be similar, but this result does not represent the geometric differences in the distribution that we can see in each one of the samples due to the non--homogeneous distribution. To characterize the differences between samples we have to consider other parameters to make comparisons. One parameter that can give us appropriate results is the intrinsic dispersion of the distribution because this parameter can characterize the shape of this distribution. This result is similar to that found in \citet{nig10} and named the ``geometrical effect", where it was shown that a good parameter to analyze structural properties of galaxies is the intrinsic dispersion of the distribution of galaxies on the plane that involves the parameters of interest. In this paper we define the intrinsic dispersion as the standard deviation of the density of galaxies at quasi-constant mass difference and/or the standard deviation of the mass difference at quasi-constant density of galaxies.

   \begin{figure}



\includegraphics[width=0.95\columnwidth]{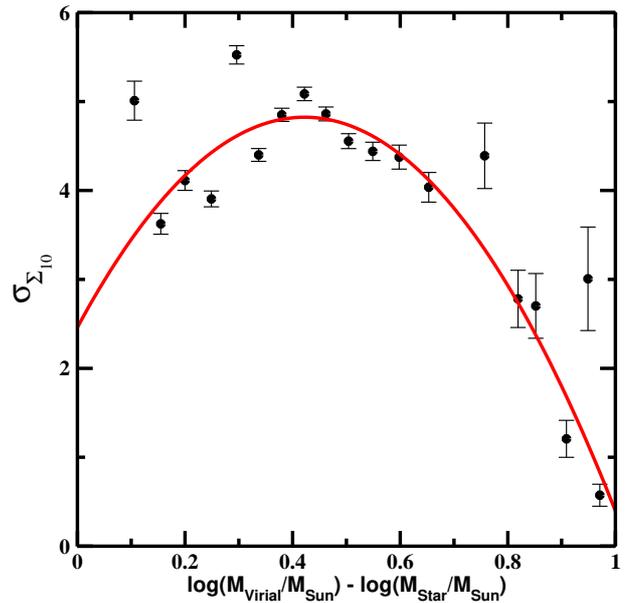}

         \caption{Intrinsic dispersion of the density of galaxies distribution (black dots) as a function of the logarithmic difference between dynamical and stellar mass (quasi--constant mass) inside $r_{e}$ for the spectroscopic-homogeneous--SDSS sample (Kroupa--IMF stellar mass) considering the tenth nearest neighbour. The red continuous line is a second degree polynomial least square fit to the data.}

         \label{FigVibStab6}

   \end{figure}

   \begin{figure}
   \begin{center}

      \includegraphics[angle=0,width=0.95\columnwidth]{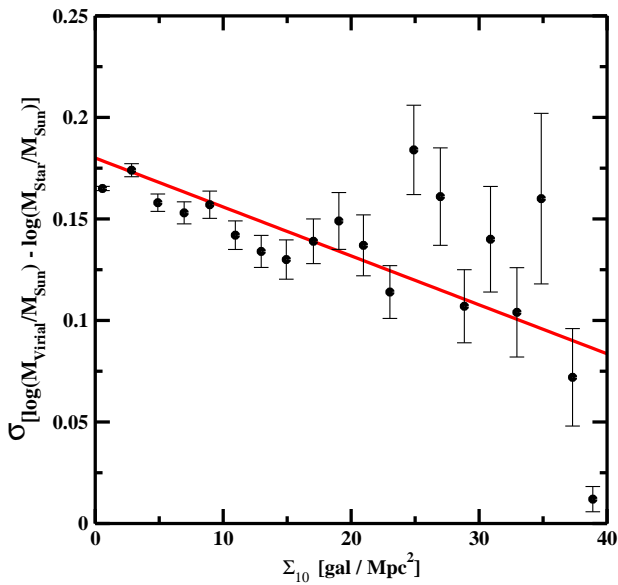}

         \caption{Intrinsic dispersion of the logarithmic difference between dynamical and stellar mass (black dots) inside $r_{e}$ as function of density of galaxies (quasi-constant density) for the spectroscopic--homogeneous--SDSS sample (Kroupa--IMF stellar mass) considering the tenth nearest neighbour. The red continuous line is a first degree polynomial least square fit to the data.}

         \label{FigVibStab7}
         \end{center}
   \end{figure}

Due to the similarity of the distribution of log$({\bf M_{Virial}/{\bf M_{Sun}}})$ -- log$({\bf M_{Star}/{\bf M_{Sun}}})$ as a function of the density of galaxies for the photometric and spectroscopic samples which consider different nearest neighbours, and to diminish biases due to incompleteness of the total and morphological SDSS samples, from here onwards we only show the analysis for the spectroscopic--homogeneous--SDSS sample considering the tenth nearest neighbour.

In subsequent paragraphs we use the intrinsic dispersion method to analyze the density distribution for the spectroscopic--homogeneous-SDSS sample considering the tenth nearest neighbour. The term quasi-constant mass in this context means mass intervals in the logarithm of width equal to 0.05 dex. The term quasi-constant density means density intervals of width equal to 2 gal Mpc$^{-2}$.

In Fig. \ref{FigVibStab6} we show the intrinsic dispersion of the density of galaxies at quasi-constant log$({\bf M_{Virial}/{\bf M_{Sun}}})$ -- log$({\bf M_{Star}/{\bf M_{Sun}}})$ (black dots). The red continuous line is a second degree polynomial least square fit to the data. The equation of the fit is as follows:

\begin{equation}  \label{eq:eq5}
{\large {\sigma_{\Sigma_{10}}}} = -13.3 X^{2} + 11.2 X + 2.5,
\end{equation}

where {${\sigma_{\Sigma_{10}}}$} is the intrinsic dispersion of the density of galaxies considering the tenth nearest neighbour and X = log$({\bf M_{Virial}/{\bf M_{Sun}}})$ -- log$({\bf M_{Star}/{\bf M_{Sun}}})$.

From this fit we obtain that the maximum density corresponds to approximately 0.4 in log$({\bf M_{Virial}/{\bf M_{Sun}}})$ -- log$({\bf M_{Star}/{\bf M_{Sun}}})$. Using linear values we find that the maximum of the density distribution corresponds to a mass difference of approximately 60\% of the virial mass.

In Fig. \ref{FigVibStab7} we show the intrinsic dispersion of the log$({\bf M_{Virial}/{\bf M_{Sun}}})$ -- log$({\bf M_{Star}/{\bf M_{Sun}}})$ at quasi-constant density of galaxies (black dots). The red continuous line is a first degree polynomial least square fit to the data. The equation of the fit is as follows:

\begin{equation}  \label{eq:eq6}
{\large {\sigma_{M}}} = -0.0024 \Sigma_{10} + 0.180,
\end{equation}

where {${\sigma_{M}}$} is the intrinsic dispersion of log$({\bf M_{Virial}/{\bf M_{Sun}}})$ -- log$({\bf M_{Star}/{\bf M_{Sun}}})$ and $\Sigma_{10}$ is the density of galaxies considering the tenth nearest neighbour.

If we consider that the difference between virial and stellar mass is due to dark matter and/or a non-universal IMF, we can say that ETGs in the lowest density regions have the widest dark matter range. This range of dark matter content decreases while the density increases (see Fig. \ref{FigVibStab7}). The amount of dark matter inside ETGs in the most dense environments and in the redshift range 0.0024 $\leq\;z\;\leq$ 0.35 would be 55\% - 65\% of the virial mass. In the case of the homogeneous-SDSS sample (0.04 $\leq\;z\;\leq$ 0.08)  the amount of dark matter in dense environments would be less than approximately 60\% of the virial mass.

It is important to mention that the previously discussed results were found using the Kroupa--IMF stellar mass, however we have found similar results for the de Vaucouleurs Salpeter--IMF and S\'ersic Salpeter--IMF samples. In table 2 we show the mean values of the difference between virial and stellar mass (\textless log$({\bf M_{virial}/{\bf M_{Sun}}})$ -- log$({\bf M_{*}/{\bf M_{Sun}}})$ \textgreater) and their intrinsic dispersion ({$\sigma_{M}$}) for the total, morphological and homogeneous samples considering both Kroupa and Salpeter IMF and both de Vaucouleurs and S\'ersic brightness profiles. The analysis of the de Vaucouleurs Salpeter-IMF and the S\'ersic Salpeter--IMF samples shows that the distribution of galaxies is similar to that found for the Kroupa--IMF samples. This finding indicates that our results are approximately independent not only of the nearest neighbour, but also of the IMF and the brightness profile. There are two differences between the Kroupa and Salpeter IMF samples that do not change our general findings (see Tab \ref{tab2}). The first one is that the average DM dispersion of the S\'ersic-Salpeter-IMF samples ({$\sigma_{M}\sim 0.12$}) is less than the dispersion of the Kroupa IMF ({$\sigma_{M}\sim 0.16$}) and the de Vaucouleurs--Salpeter IMF ({$\sigma_{M}\sim 0.16$}) samples. This finding agrees with the one obtained by \citet{tay10} that find that the dynamical and stellar masses correlate best when the structure of the galaxy is taken into account. The second difference is that there is a small shift (approximately 10\%) towards higher amounts of dark matter for the Salpeter IMF samples with respect to the Kroupa samples. Taking into account this shift due to the IMF, our final results indicate that the amount of DM inside ETGs in the most dense environments and in the redshift range 0.0024 $\leq\;z\;\leq$ 0.35 would be 55\% - 75\% of the dynamical mass. However, the accurate value depends on the impact of the IMF on the stellar mass estimation. Therefore,  this result would be an upper limit for the DM because we have to take into account that the IMF depends on mass. The real value of the DM inside ETGs in dense environments would be less than 55\% - 75\% of the virial mass. In the case of the less biased and most complete sample in mass --the homogeneous--SDSS sample (0.04 $\leq\;z\;\leq$ 0.08)-- we can say that the amount of DM in dense environments would be less than approximately 60\% - 65\% of the virial mass.

Here it is interesting to note that previous results indicate a small shift towards higher dark matter for the Salpeter IMF samples with respect to the Kroupa IMF samples, however in the literature we can find that the Salpeter IMF gives 50\% more stellar mass than the Kroupa IMF. This apparent contradiction can be explained considering that in this work a ``diet" Salpeter IMF was used, this ``diet" Salpeter IMF produces 30 percent less stellar mass than a normal Salpeter IMF (\citealt{bel03}). Taking this point into account, the small shift of the DM calculated using a ``diet" Salpeter IMF is not significant and this DM amount agrees, within the errors, with that obtained with a Kroupa IMF.

\begin{table*}

\renewcommand{\footnoterule}{}  

\caption{Mean values of the difference between virial and stellar mass (\textless log$({\bf M_{virial}/{\bf M_{Sun}}})$ -- log$({\bf M_{*}/{\bf M_{Sun}}})$ \textgreater) and their intrinsic dispersion ({$\sigma_{M}$}) for the total, morphological and homogeneous samples considering different IMF (Kroupa, Salpeter) and different brightness profiles (de Vaucouleurs, S\'ersic).}

\scalebox{0.45}{
\begin{tabular}{|l|l|c|c|c|c|}

\hline
                           &                             &                 &                                                      & &  \\

              &  &{\Huge \textless log$({\bf M_{virial}/{\bf M_{Sun}}})$ - log$({\bf M_{*}/{\bf M_{Sun}}})$ \textgreater } &{\Huge {$\sigma_{M}$}}                          & {\Huge }   \\
                           &                             &                 &                                                          \\
\hline

        &                             &                 &                                                         \\

{\Huge Total sample}        &   & &    &               \\
                           &                             &                 &                                                         \\
\hline
                           &                             &                 &       &                                                   \\

  &{\Huge Kroupa IMF }  &{\Huge 0.416} &{\Huge 0.158 $\pm$ 0.118} &  \\
      &     &                             &                 &            \\
        &     &                             &                 &             \\
                         
 &{\Huge de Vaucouleurs Salpeter-IMF}   &{\Huge 0.493} & {\Huge 0.155 $\pm$ 0.115} & \\

      &     &                             &                 &            \\
        &     &                             &                 &             \\

 &{\Huge S\'ersic Salpeter-IMF}   &{\Huge 0.484} & {\Huge 0.113 $\pm$ 0.084} & \\

\hline

     &                             &                 &                                                        \\

&   & &    &               \\
{\Huge Morphological sample}        &   & &    &               \\
                           &                             &                 &                                                         \\
\hline
                           &                             &                 &       &                                                   \\

 &{\Huge Kroupa IMF }  &{\Huge 0.421} &{\Huge 0.151 $\pm$ 0.113} &  \\
      &     &                             &                 &            \\
        &     &                             &                 &             \\
                         
 &{\Huge de Vaucouleurs Salpeter-IMF}   &{\Huge 0.486} & {\Huge 0.145 $\pm$ 0.108} & \\

      &     &                             &                 &            \\
        &     &                             &                 &             \\

 &{\Huge S\'ersic Salpeter-IMF}   &{\Huge 0.461} & {\Huge 0.116 $\pm$ 0.084} & \\

\hline
                           &                             &                 &                                         &               &  \\

&   & &    &               \\
{\Huge Homogeneous sample}        &   & &    &               \\
                           &                             &                 &                                                         \\
\hline
                           &                             &                 &       &                                                   \\

  &{\Huge Kroupa IMF }  &{\Huge 0.391} &{\Huge 0.165 $\pm$ 0.123} &  \\
      &     &                             &                 &            \\
        &     &                             &                 &             \\
                         
 &{\Huge de Vaucouleurs Salpeter-IMF}   &{\Huge 0.504} & {\Huge 0.162 $\pm$ 0.121} & \\

      &     &                             &                 &            \\
        &     &                             &                 &             \\

 &{\Huge S\'ersic Salpeter-IMF}   &{\Huge 0.486} & {\Huge 0.117 $\pm$ 0.087} & \\

\hline

\end{tabular}}
\label{tab2}

\end{table*}

\section{Systematics}
\label{sec:systematics}

There are some systematic uncertainties that have to be taken into account in our mass estimation. In the following we show the uncertainties associated with the different variables and the final error in our DM estimation.

To calculate the de Vaucouleurs Salpeter-IMF stellar mass we used the model--parametric photometric information, as well as its associated errors, from the SDSS-DR9 (\citealt{nig15}). The S\'ersic Salpeter--IMF stellar mass were obtained using the Petrosian magnitude and Petrosian radius and their associated errors from the SDSS-DR9
\citep{nig15}. The Kroupa--IMF stellar mass were obtained using the model--parametric photometric information, as well as its associated errors, from the SDSS-DR9 (\citealt{kau03}). In the case of the virial mass, besides the effective radius, the velocity dispersion is required, which was calculated inside the radius subtended by the SDSS fibre and corrected to an aperture independent of both the distance and
instrument used for the observations. All the above parameters were corrected for various sources of bias (see section \ref{sec:sample}) and the errors were obtained considering the rules of error propagation. The errors of the photometric and spectroscopic parameters from the SDSS--DR9 are the same we have used in previous papers, for full details see \citet{nig15}. However, taking into account different sources of systematics, several papers from the literature \citep{ems04,che10,cap12,cap13} show that the errors from the SDSS--DR9 may be underestimated. Considering the above-mentioned studies, in this work we adopt the following mean errors: 0.1 mag, 10\%, and 5\% for the magnitude, effective radius, and velocity dispersion, respectively. Taking into account this information the mean uncertainties for the stellar and virial mass are approximately 15\% and 12\%, respectively.

In addition in this paper we have considered that the scale factor K in the virial relation (see equation \ref{eq:eq2}) depends only on the surface-brightness profile and that the DM follows the same density profile as the stellar component which, according to some authors \citep{cio96,koo06,tho11,tor12b,tor18}, is not appropriate for compact and massive galaxies, respectively. Following \citet{tor12b,tor18} we find that the above-mentioned considerations lead to an error on dark matter up to 7\% and 20\%, respectively. The error in the scale factor K includes and error of approximately 4\% for the S\'ersic n index.

We also have to consider that the velocity dispersion of the ETGs is not necessarily isotropic and that these galaxies could have different degrees of rotation. Using cosmological hydrodynamical simulations of the local group of galaxies, \citet{cam17} find that the dynamical mass obtained with estimators such as equation \ref{eq:eq2} of this paper (see equation 18 of \citealt{cam17} for comparison), without considering that different environment, shape, rotation and/or velocity dispersion anisotropies, could render it possible to recover the masses of dispersion-dominated systems with little systematic bias but with a scatter of 20\% approximately.

If we consider all this information the final uncertainty on the DM estimation increases approximately to 34\%.

\section{Discussion}
\label{sec:discussion}

Over the last few years, different authors have studied the dependence of the dynamical and stellar mass inside the effective radius of ETGs on several variables such as dynamical (or stellar) mass and redshift. The vast majority of these studies show that log$({\bf M_{Virial}/{\bf M_{Sun}}})$ -- log$({\bf M_{Star}/{\bf M_{Sun}}})$ increases with galaxy mass and decreases with redshift. The origin of this trend remains controversial. It is still under debate whether this phenomenon is driven by DM and/or a non-universal IMF (\citealt{dut13,tho11,weg12,con13,nig16}). The above-mentioned results are important but do not cover all the variables that can affect the difference between dynamical and stellar mass. There are other variables (e. g., environment) that it is necessary to analyse to have a complete picture of the above-mentioned ratio. However, there are only a few works in the literature where these variables are investigated. This paper is devoted to the study of one of these variables; the environment. To analyse log$({\bf M_{Virial}/{\bf M_{Sun}}})$ -- log$({\bf M_{Star}/{\bf M_{Sun}}})$ as function of the environment, we have considered the density of galaxies using the definition of the nearest neighbour and different considerations to avoid, as far as possible, systematic biases (see section \ref{sec:sec3.3}).
Our results show that the difference between dynamical and stellar mass depends on density and that the ETGs in low density environments span a wider range than ETGs in dense environments. We also find that the difference between dynamical and stellar mass inside ETGs in the most dense environments is approximately 55\% - 75\% of the dynamical mass. These results are interesting and have to be interpreted considering the source of biases and systematics that are affecting our mass estimations. The above-mentioned systematics indicate that our difference between dynamical and stellar mass has an error of up to 34\% (see section \ref{sec:systematics}). We also have to consider that this difference between dynamical and stellar mass could originate due to DM and/or variations of the IMF of early type systems, being 55\% - 75\% of the dynamical mass an approximate upper limit for the amount of DM, the accurate value depends on the impact of the IMF on the stellar mass estimation.

The above-mentioned results can be compared with other results from the literature, however there is only one paper with which this comparison can be done directly. \cite{tor12b} performed an analysis of the amount of DM inside $r_{e}$ in ETGs, they used approximately 4500 ETGs from the SDSS and supplemented their data with $YJHK$ photometry from the DR2 of the UKIDSS-Large Area Survey. They found that the central DM content of ETGs did not depend significantly on the environment where galaxies resided, with group and field ETGs having similar DM trends. This result agrees with the result found in section \ref{sec:stellar&virial} of this paper where it is shown that if we consider low and high density environments the amount of DM inside ETGs appears to be similar. However, it is an artifact. As was also discussed in section \ref{sec:stellar&virial}, this similarity is due to the way in which the comparison is done (geometrical effect). The distribution of DM as a function of density has a high intrinsic dispersion and to obtain good results it is necessary to analyse this intrinsic dispersion at quasi-constant density and/or DM. When this kind of analysis is done we find that ETGs in low density environments span a wider DM range than ETGs in dense environments and that the DM inside ETGs in the most dense environments is less than approximately 55\% - 75\% of the dynamical mass as mentioned above. On the other hand \citet{cor17} find that ETGs in low density environments have a lower content of halo DM with respect to ETGs in high density environments. If we consider that the behaviour of the DM in the central part of the halo ($r_{e}$) is similar to the behaviour of the DM in the entire halo of the ETGs, the results of \citet{cor17} agree with our results. Therefore, there are galaxies in low density environment that have less DM than galaxies in high density environments as shown in Fig. \ref{FigVibStab4}.

Here it is important to mention that a recent paper from the literature (\citealt{cam17}), using cosmological hydrodynamical simulations of the local group of galaxies, find that the dynamical masses calculated with estimators such as equation \ref{eq:eq2} do not show significant differences, on average, due to the environment where the galaxies reside. The only difference that they find is that field galaxies show bigger scatter (approximately 5\%, see figure 12 of \citealt{cam17}) in the dynamical mass estimation than galaxies in denser environments. They find that this behaviour may be related to the fact that, on average, the galaxies in denser environments are closer to spherical symmetry than those in the field, they also show hints that galaxies in denser environments tend to be more strongly supported by dispersion and tend to have stellar velocity dispersions that are closer to isotropic, compared with field galaxies. The scatter, due to environment, on the virial mass estimation obtained with equation \ref{eq:eq2} is small compared with the average scatter of the dark matter found in this work ({$\sigma_{M} \sim 30\%$}). Besides, if they consider other sources of bias such as shape, rotation and velocity dispersion anisotropies the total scatter on dynamical mass due to the estimator is approximately 20\%. This total scatter is significantly smaller than the scatter on DM found in this work, so the scatter due to the dynamical mass estimator cannot entirely explain our findings.

Considering these results, it is necessary to undertake better and more thorough studies improving all possible sources of bias and systematics behind the behaviour of differences between dynamical and stellar mass. It is also important to analyse the behaviour of the difference between dynamical and stellar mass in a wide range of wavelengths. We shall address all these issues in a forthcoming paper.

\section{Conclusions}
\label{sec:conclusions}

The analysis of the distribution of log$({\bf M_{Virial}/{\bf M_{Sun}}})$ -- log$({\bf M_{Star}/{\bf M_{Sun}}})$ vs. density of galaxies for several samples of ETGs from the SDSS-DR9 in the redshift range 0.0024 $\leq\;z\;\leq$ 0.35 and considering that this difference is due to DM and/or a non-universal IMF, has yielded the following results:

    i)   The DM as a function of the environment does not have a homogeneous distribution.

    ii)  The amount of DM inside ETGs depends on the environment.

    iii) ETGs in low density environments span a wider DM range than ETGs in dense environments.

    iv) The amount of DM inside ETGs in the most dense environments and in the redshift range 0.0024 $\leq\;z\;\leq$ 0.35 is approximately 55\% - 75\% of the dynamical mass

    v)  This amount of DM is an upper limit, the accurate value depends on the impact of the IMF on the stellar mass estimation.

    vi)   In the case of the less biased and more complete (in mass) ETGs sample -the homogeneous sample (0.04 $\leq\;z\;\leq$ 0.08)- we can say that the amount of DM in dense environments will be less than approximately 60\% - 65\% of the dynamical mass.

\section*{Acknowledgements}

We thank the Instituto de Astronom\'{\i}a y Meteorolog\'{\i}a (UdG, M\'exico) and Instituto de Astronom\'{\i}a (UNAM, M\'exico) for all the facilities provided for the realisation of this project. A. Nigoche-Netro and G. Ramos-Larios acknowledge support from CONACyT and PRODEP (M\'exico). E. de la Fuente acknowledges financial support from Laboratorio Nacional de CONACyT `HAWC' de rayos gamma. A. Ruelas-Mayorga thanks the Direcci\'on General de Asuntos del Personal Acad\'emico, DGAPA at UNAM for financial support under projects number PAPIIT IN103813 and PAPIIT IN102617. P. Lagos is supported in the form of work contract (DL 57/2016/CP1364/CT0010) funded by national funds through Funda\c{c}\~ao para a Ci\^encia e Tecnologia (FCT), Portugal. J. Mendez-Abreu acknowledge support from the Spanish Ministerio de Economia y Competitividad (MINECO) by the grant AYA2017-83204-P. Last but not least, we thank and acknowledge the comments made by an anonymous referee, which improved greatly the presentation of this paper.

\bsp	
\label{lastpage}
\end{document}